# Examining the Spin Structure of Altermagnetic Candidate MnTe Grown with Near Ideal Stoichiometry

Qihua Zhang[1*], Christopher J. Jensen[2], Alexander J. Grutter[2], Sandra Santhosh[3], William D. Ratcliff[2,4,5], Julie A. Borchers[2], Thomas W. Heitmann[6], Narendirakumar Narayanan[6], Timothy R. Charlton[7], Mingyu Yu[8], Ke Wang[9], Wesley Auker[9], Nitin Samarth[1, 3, 9,10,11], Stephanie Law[1, 3, 9,10, 12]*

[1]Two Dimensional Crystal Consortium Materials Innovation Platform, The Pennsylvania State University, University Park, Pennsylvania 16802, USA

[2]NIST Center for Neutron Research, NCNR, National Institute of Standards and Technology, Gaithersburg, Maryland 20899, USA

[3]Department of Physics, The Pennsylvania State University, University Park, Pennsylvania 16802, USA

[4]Department of Physics, University of Maryland, College Park, Maryland 20742, USA

[5]Department of Materials Science and Engineering, University of Maryland, College Park, Maryland 20742, USA

[6]University of Missouri Research Reactor (MURR), University of Missouri, Columbia, Missouri 65211, USA

[7]Neutron Scattering Division, Oak Ridge National Laboratory, Oak Ridge, Tennessee 37831, USA

[8]Department of Materials Science and Engineering, University of Delaware, 201 Dupont Hall, 127 The Green, Newark, Delaware 19716, USA

[9]Materials Research Institute, The Pennsylvania State University, University Park, Pennsylvania 16802 USA

[10]Department of Materials Science and Engineering, The Pennsylvania State University, University Park, Pennsylvania 16802 USA

[11]Argonne National Lab, Chicago, Illinois 60439 USA

[12]Institute of Energy and the Environment, The Pennsylvania State University, University Park, Pennsylvania 16802 USA

ABSTRACT Altermagnets are a recently-discovered class of materials with magnetic ordering that have a zero net magnetization and a momentum-dependent spin splitting in their band structure, arising from a collinear spin arrangement with alternating polarizations in the crystal lattice. The nickeline-structured manganese telluride (α-MnTe) is an attractive altermagnet candidate due to its predicted large spin splitting energy and a transition temperature near 300K. In this work, we present a thorough investigation of the spin structure of α-MnTe thin films grown by molecular beam epitaxy with very high crystal quality and low residual magnetization. The epitaxial α-MnTe films have a full-width-at-half-maximum of 0.1° as measured by x-ray-diffraction rocking curves and a root-mean-square roughness below 1 nm. Neutron diffraction measurements confirm the antiferromagnetic order in the α-MnTe film and show a Néel temperature of 307 K. Polarized neutron reflectometry detects a vanishingly small net magnetization which may be confined to the MnTe/InP interface, highlighting the near-ideal stoichiometry in the sample. *In vacuo* angle resolved photoemission spectroscopy reveals that the bulk band spectrum of the MnTe films is consistent with the weak altermagnetic order as theoretically predicted and observed for the high symmetry nodal plane in the center of the Brillouin zone. This study establishes optimized growth conditions for the synthesis of stoichiometric α-MnTe thin films which exhibit exceptional structural and magnetic ordering, thereby providing a robust platform for the precise characterization of their altermagnetic properties.

## 1. Introduction

Recently, a new class of magnetic materials known as "altermagnets" has attracted significant research interest since their magnetic properties are unique compared to ferromagnets (FMs) and antiferromagnets (AFMs).[1-10] Altermagnets feature a net zero magnetization in their bulk form similar to a traditional AFM, and they exhibit a collinear spin arrangement with alternating spin polarizations in their crystal lattice. However, they also possess an anisotropic spin splitting in their electronic band structure, a feature akin to FMs.[11] The unique properties of altermagnets give rise to a myriad of potential applications in a range of fields including spintronic devices, quantum computing, high-density data storage, and so on. The Nickeline-structured α-MnTe is one of the most attractive altermagnetic materials among all the theorized candidates due to its near room temperature Néel temperature and large predicted spin splitting energy.[12-15] Historically, α-MnTe was classified as an A-type antiferromagnet with a Néel temperature ($T_N$) of ~310 K.[13, 16-21] The spins are ferromagnetically aligned within each basal plane which are oriented antiparallel to each other along the [0 0 0 $l$] c-axis direction. Prior reports confirm that the Néel vector orientation is parallel to the basal plane with negligibly small canting along the [0 0 0 $l$] direction.[16-18] It was not until very recently that Šmejkal et al. used *ab initio* calculations to demonstrate that altermagnetic α-MnTe features a *g*-wave anisotropy and a large spin splitting energy of 1.1 eV in the valence band, which is one of the largest predicted among all altermagnetic semiconductor candidates.[2]

With its established altermagnetic candidacy, substantial work has thus been focused on reassessing MnTe to measure its altermagnetic characteristics.[22-32] A number of reports examining the altermagnetic properties of α-MnTe thin films using angle-resolved photoemission spectroscopy (ARPES), transport measurements, and x-ray absorption spectroscopy have been reported in the past. By combining ARPES data with disordered local moment (DLM) calculations, splitting of the valence band that disappeared as the temperature rose above the Néel temperature was observed, suggesting a connection between the valence band splitting and altermagnetic spin splitting order.[26] Furthermore, by combining x-ray magnetic linear dichroism (XMLD) and x-ray magnetic circular dichroism (XMCD) with photoelectron emission microscopy (PEEM) measurements, the mapping of altermagnetic order vector was revealed, which showed controlled formations of altermagnetic configurations of diverse sizes, spanning from nanoscale vortices to microscale domains.[23, 27]

However, the detailed magnetic and spin structure of pure α-MnTe thin films to date remains unclear and may be influenced by strain and/or stoichiometry deviations. Other studies, however, indicate that the bulk antiferromagnetic order is preserved in film form.[18,19,33] Specifically, it is unknown if defect-free films have an intrinsic canted magnetic moment that gives rise to a small net ferromagnetic magnetization. Although Chicote *et al.* investigated the spin structure and spin asymmetry of α-MnTe film using polarized neutron reflectometry (PNR), the film showed a nonuniform stoichiometric ratio as suggested by a graded nuclear scattering length density ($\rho$) in the MnTe region as well as a surface oxide layer of 6-7 nm thick.[33] A significant net ferromagnetic moment (i.e., depth-averaged magnetization of 24 kA/m through the film) was indicated by the spin asymmetry between the reflectivity polarization states, and the magnetic scattering length density ($\rho_M$) modeled in the structures highlights the difficulty in classifying the origin of the net magnetization in the MnTe film. Other past reports also acknowledge the existence of substantial net magnetization in the α-MnTe films and attribute it to either a high density of structural defects or intermixing in the film/substrate interfacial region.[18, 21, 34-37] A single-phase α-MnTe thin film with careful control of stoichiometric ratio and pristine surface is thus needed to fully understand its spin structure and altermagnetic properties.

Here, we present detailed investigations of the magnetic spin structure of high-quality epitaxial MnTe thin films, with carefully controlled stoichiometry, grown by molecular beam epitaxy (MBE) with a focus on determining the existence of spin canting. By varying the growth parameters including substrate temperature and Te/Mn flux ratio (FR), we obtain α-MnTe thin films grown on InP (111)A substrates with high structural quality, as characterized by a surface roughness Rq ~ 1 nm (over 2 × 2 μm$^2$) and a narrow full-width-at-half-maximum (FWHM) of 0.1° and 0.3° at the MnTe 0004 and 10$\bar{1}$5 x-ray-diffraction (XRD) rocking curves. We use transmission electron microscopy (TEM) and high angle neutron diffraction to confirm the Nickeline crystal structure and the Néel temperature of the α-MnTe film. Using PNR measurements to examine the ferromagnetic component of the MnTe film, we find an average magnetization throughout the film of 2.6 kA/m +/- 0.5 kA/m, which is more than 2 – 4 times smaller than previously reported PNR works on α-MnTe films.[19, 33] This improvement in net magnetization likely stems from either a reduced number of defects in the MnTe film or an improved MnTe/InP interfacial region, which denotes a near ideal film stoichiometry. We further use *in vacuo* ARPES measurements to probe the band structure and find that it corresponds well with previous reports.[22, 24] This paper establishes growth conditions and protocols that allow the synthesis of stoichiometric α-MnTe films of high structural and magnetic order, thus providing a platform for careful evaluation of altermagnetic properties that are not influenced by extrinsic factors such as inhomogeneous ferromagnet phases.

## 2. Experimental Procedures

*MBE growth*

In this report, all MnTe films were grown in a DCA Instruments R450 MBE system (instrument details at 10.60551/gqq8-yj90) with a base pressure of 5 × 10$^{-10}$ Torr. InP (111)A wafers diced into 1 × 1 cm$^2$ pieces were used as substrates. Standard solvent cleaning procedures (acetone and isopropyl alcohol ultrasonic bath) and thermal outgassing at 200 °C for 2 hours in the load lock chamber were performed on the substrates prior to sample growth. After outgassing in the load lock, the substrate was heated to 650 °C at a ramp rate of 30 °C/min in the MBE system and annealed for 10 min with a constant supply of tellurium at a flux of 2.3 × 10$^{14}$ cm$^{-2}$ s$^{-1}$ to desorb the surface oxide.[38] The substrate temperature was measured by a non-contact thermocouple mounted behind the substrate. A STAIB reflection high electron diffraction (RHEED) electron gun and kSA 400 analytical RHEED software were used to monitor the surface *in situ* during the substrate anneal and sample growth. Te and Mn fluxes of >5N purity were supplied in standard Knudsen cells. A ColnaTec quartz crystal microbalance (QCM) operating at 6 MHz was used for flux calibrations at the growth position. For the growth optimization studies, a consistent Mn flux of 7.6 × 10$^{13}$ cm$^{-2}$ s$^{-1}$ was used, and growth durations were fixed at 20 minutes with a target film thickness of ~ 30 nm. The Te flux was varied from 7.6 × 10$^{13}$ cm$^{-2}$ s$^{-1}$ to 3.1 × 10$^{14}$ cm$^{-2}$ s$^{-1}$, while the growth temperature varied from 200 °C to 500 °C. After the film deposition was completed, the sample was annealed for 2 minutes at the growth temperature with the Te shutter open and then ramped down to room temperature at a rate of 50 °C/min. For samples used for the PNR experiment, a 5 nm amorphous Te layer was deposited on the MnTe sample in the MBE system at room temperature with a Te flux of 1 × 10$^{14}$ cm$^{-2}$/s to prevent oxidation of the MnTe layer.

*Angle Resolved Photoemission Spectroscopy (ARPES)*

ARPES measurements were performed on a 35 nm thick MnTe film using a Helium lamp with a photon energy of 21.2 eV. The detection of the photoemitted electrons were done using a Scienta Omicron DA 30L analyzer with an energy resolution of 6 meV. The samples were transferred from the growth chamber to an ultra-high vacuum ARPES chamber using a vacuum suitcase with a base pressure less than 5x10$^{-8}$ mbar. The samples were measured at 298 K and 77 K.

*Structural and surface analysis of MnTe Films*

High resolution XRD measurements were conducted using a Malvern PANalytical 4-circle X'Pert[3] system equipped with a hybrid 2-bounce asymmetric Ge(220) monochromator and a PIXcel 3D detector. The surface morphology was examined in a Bruker Dimension Icon atomic force microscope (AFM) with a lateral resolution of 512 pixels/line. A SCANASYST-AIR probe was used for AFM imaging. Thin cross-sectional transmission electron microscope (TEM) specimens were prepared by using focused ion beam (FIB, FEI Helios 660) lift-out technique. A thick protective amorphous carbon layer was deposited over the region of interest prior to FIB, and a 2 kV final cleaning was used to remove the ion beam damage the sample surface. Then the cross-sectional microstructures of the films were observed by FEI Titan3 G2 double aberration-corrected microscope at 300 kV. All scanning transmission electron microscope (STEM) images were collected by using a high-angle annular dark field (HAADF) detector which had a collection angle of 52-253 mrad. EDS elemental maps of the sample were collected by using a SuperX EDS system under STEM mode.

*High Angle Neutron Diffraction*

Elastic neutron scattering measurements on 35 nm and 100 nm MnTe films on InP (111) substrates (lateral area of 1 cm$^2$) were performed in zero applied magnetic field at the University of Missouri Research Reactor (MURR) using the triple-axis spectrometer TRIAX. A pyrolytic graphite (PG) monochromator was used to select an incident beam energy of 14.7 meV, and a PG analyzer after the sample was also set to 14.7 meV to reduce the background. Higher wavelength harmonics were reduced with PG filters positioned in the incident and scattered beams. The beam divergence was defined using 60′-60′-80′-80′ collimators positioned before the monochromator, between the monochromator and sample, between the sample and analyzer, and between the analyzer and detector, respectively. The sample was sealed in a $^4$He environment within an aluminum can and subsequently mounted on the cold tip of an Advanced Research Systems closed-cycle refrigerator. The films were aligned with the [0 0 0 *l*] MnTe growth axis parallel to the wavevector Q direction, and θ/2θ scans were performed through the 0 0 0 1 antiferromagnetic reflection at different temperatures upon heating the sample. These data were fit with Gaussians, and the integrated peak intensities were plotted as a function of temperature to produce antiferromagnetic order parameters.

*Polarized Neutron Reflectometry*

PNR experiments were conducted on the MAGREF reflectometer at the Spallation Neutron Source at Oak Ridge National Laboratory on Te (5 nm)/MnTe (35 nm)/InP films with a lateral area of 1 cm$^2$ at 20 K and 100 K in a 1 T field to further explore their structural and magnetic properties. For these measurements, a spin-polarized neutron beam with a wavelength band λ (0.354 – 0.942 nm) was produced using a two-mirror v-design transmission supermirror polarizer. The spin state of the incident neutrons was selected to be up (+) or down (-) with a spin flipper. Two samples were measured simultaneously, and the scattered beam was measured at four different values of the incident angle θ to cover an extended Q range for each. The spin-dependent neutron reflectivities, $R^+$ and $R^-$, were then obtained as a function of the wavevector transfer Q in the direction normal to the film surface. Only polarized spin-up ($R^+$) and spin-down ($R^-$) cross-sections of the reflectivity were collected since a 1 T field is enough to saturate any parasitic moment in MnTe at the measurement temperatures.

PNR probes the nuclear and magnetic scattering length densities (ρ and $ρ_M$, respectively), the latter of which is proportional to the magnitude of the ferromagnetic component of the film magnetization aligned parallel to the applied magnetic field. These components can be extracted through modeling and fitting of the measured reflectivity data, and the ρ and $ρ_M$ depth profiles can be reconstructed as a function distance from the film-substrate interface, Z, normal to a sample surface. Data reduction for the experiments was completed using the QuickNXS extraction software, and fitting and uncertainty analysis were performed using the Refl1d software packages.[39, 40] The values of the incident intensities and angle offset were varied independently in fits of the reflectivity data obtained at different values of the incident angle θ.

In typical PNR experiments, there is degeneracy in the models used for fitting, where multiple depth profiles yield similar $\chi^2$ values. As a result, a unique reconstruction of the sample structure in real space is not possible. This issue requires limiting potential models to those that are consistent with the known physical parameters of the system. During the fitting process we started from the simplest model, non-magnetic MnTe and a uniform ρ for all layers, and then we produced a series of increasingly complex models to determine the point of under or over parameterization. To further improve the models and uncertainties, fitting was performed on data collected in 1 T at 20 K and 100 K simultaneously, which allowed parameters to be constrained. Our models used the assumptions that the sample structure did not change between the two temperatures, except for the surface layer ("SL" in Fig. 5) – representative of ice and/or condensate accumulation while cooling. Parameters related to magnetism, such as $\rho_M$ and magnetic thickness, were allowed to vary between the two temperatures, as changes in superparamagnetic clusters and/or changes in magnetic anisotropy could produce differences between the two temperatures. A more detailed description of the fitting process and models representing under and over parameterization are included in the "Polarized Neutron Reflectometry" section of the Supplemental Material.

## 3. Results and Discussion

3.1 Optimization of growth conditions for α-MnTe layers

*Growth Temperature*

We first investigated the effect of growth temperature over a range of 200 – 500°C (measured by thermocouple) on the quality of MnTe films. The samples in this series were grown with a fixed growth rate of 0.35 Å/s (1 Å = 0.1 nm), a fixed Te/Mn flux ratio (FR) of 3, and a constant growth duration of 20 min. All samples have similar MnTe film thickness of 26 – 29 nm as measured by x-ray reflectivity (XRR) fitting, suggesting that the desorption rate for Mn adatoms is minimal within this wide growth temperature window. We first compared the surface morphology of Samples A – F, grown with a temperature of 200 °C (Sample A), 250 °C (Sample B), 300 °C (Sample C), 340 °C (Sample D), 400 °C (Sample E), and 500 °C (Sample F), respectively. As seen in Figure 1, triangular-like and hexagonal-like grains on all the six samples are observed, consistent with the desired hexagonal-phase MnTe. However, Sample A, grown with a temperature of 200 °C, has a number of needle-like whiskers on the surface. We suspect that these whiskers are $MnTe_2$ pyrite crystals formed randomly on the surface, owing to the low diffusion length of both Mn and Te adatoms from the low growth temperature; however, the quantity of the whiskers is insufficient to be detected by XRD scans. By increasing the growth temperature from 250°C to 500°C, one can see clear step edge formation along with increasing grain sizes due to the increase in adatom mobility and a disappearance of the whiskers. However, the surface also becomes more columnar and island-like, with RMS roughness gradually increasing from ~1.0 nm to 3.0 nm from Sample B to F. This suggests that a high growth temperature is not favorable for MnTe growth, as the high desorption rate of the Te adatoms compromises the MnTe nuclei formation during the initial stages of growth.

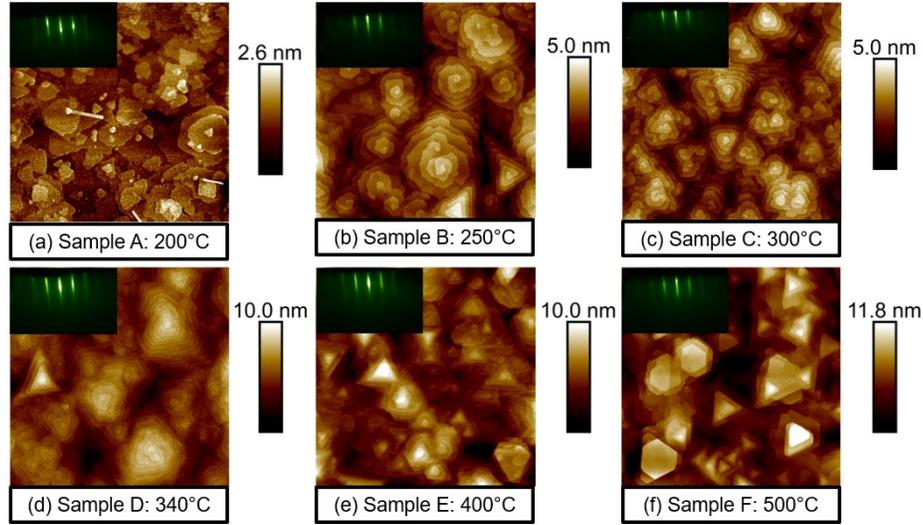

**Figure 1.** AFM images of MnTe films grown with growth temperatures of (a) 200°C, (b) 250°C, (c) 300°C, (d) 340°C, (e) 400°C, and (f) 500°C. All AFM scans are 2 × 2 µm². The inset displays the corresponding RHEED images captured immediately after growth at the growth temperature.

As both Mn and Te fluxes were kept constant in this series of samples, the increased Te desorption rate with the increased growth temperature reduces the effective Te/Mn ratio at the growth front, which in turn reduces the crystalline quality of the synthesized MnTe films. To confirm this, we carried out high resolution XRD studies of Samples A – F. Exemplary XRD scans including the 2θ-ω and the ω-scans (i.e., rocking curves) of the MnTe 0004 and $10\bar{1}5$ reflections are shown in Figure S1. Using the MnTe (0004) 2θ peak position at ~54.71°, the *c*-axis lattice constant of MnTe film in Sample A is calculated to be ~6.71 Å, which agrees very well with the bulk MnTe values reported in past studies.[12, 13, 18] We note that the FWHM values in the XRD rocking curves for both the (0004) and $(10\bar{1}5)$ reflections (0.09° and 0.21°, respectively) as presented in Figure S1(b-c) are among the lowest for heteroepitaxial chalcogenide thin films that are only ~30 nm thick; this high quality is expected due to the small lattice mismatch between MnTe and InP.[38]

Figure S2(a) plots the FWHM values of the XRD rocking curves for the MnTe 0004 symmetric reflection and the $10\bar{1}5$ asymmetric reflection as a function of growth temperature. We see that the FWHM values in both MnTe reflections increase with the increasing growth temperature, indicating an increase in threading dislocations in the film. This confirms that the high Te desorption rate at the high growth temperature degrades the crystalline quality. In summary, we find that a lower growth temperature improves the crystal quality of MnTe layers but promotes the growth of surface features such as whiskers. We therefore conclude that the optimized window for the growth temperature is 250 — 400°C.

*Te/Mn flux ratio*

Based on the results from growth temperature study, we determined that controlling the Te desorption is critical to improving the MnTe film quality as it heavily influences the dynamics at the growth front. Therefore, we further studied the effect of Te/Mn FR on the MnTe film quality. In this series, the growth temperature and growth rate were fixed at 340 °C and 0.35 Å/s, respectively. Figure 2 (a-d) displays the AFM images of MnTe layers grown with a Te/Mn FR of 1 (Sample G), 2 (Sample H), 3 (Sample D), and 4 (Sample J). Although the RMS roughness in Sample G (Te/Mn = 1) is the lowest among the four at 0.8 nm, the morphology in this sample is granular-like and non-continuous. The small, dense grains are rounded and globular, which is not consistent with the hexagonal- or triangular-like grains that are typically observed in (0001)-oriented thin films. We speculate that these grains are Mn droplets or Mn clusters, which are caused by the insufficient Te supply at the growth front. As the Te/Mn FR increases from 2 to 4 (Figure 2 b-d), the RMS roughness decreases from 2.0 nm to 1.2 nm. This effect can be understood by realizing that the increasing Te supply compensates for the high Te desorption rate at the growth temperature. This is also consistent with the increased surface roughness with increasing growth temperature as discussed earlier. We therefore determine that a high Te/Mn FR benefits the surface morphology of MnTe films. Unfortunately, in this study, the maximum Te flux we can achieve is

~ 3.1 × $10^{14}$ cm$^{-2}$/s, which is very close to the physical and thermal limitation of the Te effusion cell. We also note that maintaining a high operating temperature for Te effusion cell increases the likelihood of clogging in the effusion cell.

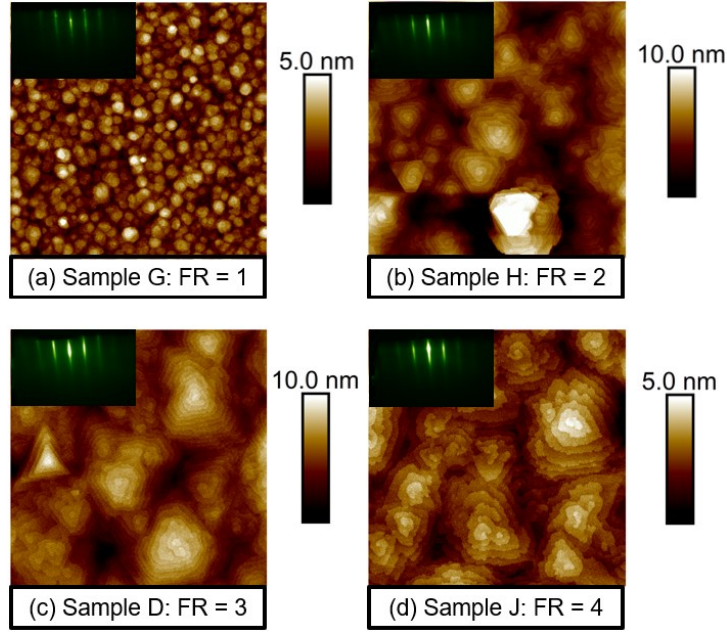

**Figure 2.** AFM images of MnTe films grown with Te/Mn flux ratios of (a) 1, (b) 2, (c) 3, and (d) 4. Sample D is repeated from Figure 1 and displayed again for comparison purposes. All AFM scans are 2 × 2 µm². The RMS roughness in the images is (a) 0.8 nm, (b) 2.0 nm, (c) 1.6 nm, and (d) 1.2 nm. The inset displays the corresponding RHEED images captured immediately after growth at the growth temperature.

In addition to the AFM scans, we can also compare the crystalline and structural quality of the MnTe films in this series with XRD. The FWHM values of the MnTe (0004) and (10$\bar{1}$5) reflections XRD rocking curves are plotted as a function of Te/Mn flux ratio in Figure S2(b). The FWHM of the MnTe (0004) rocking curves for all four samples are in the range 0.1° – 0.2°. It is again clear that a higher Te/Mn FR is beneficial to the crystalline quality of the MnTe layers, since the FWHM values of the MnTe (0004) and (10$\bar{1}$5) rocking curves both decrease with the increasing Te flux. On the other hand, oversupplying Te flux will eventually undermine the film quality either due to a decrease in the Mn adatom surface diffusion length causing cluster formations or due to the formation of additional phases such as $MnTe_2$.

3.2 Structural characterization of epitaxial MnTe films

Taking into consideration the limits of the Te effusion cell, we find an optimal growth window for growth temperature and Te/Mn flux ratio to be ~250°C – 350°C and 3, respectively, where the MnTe films have a clean surface as well as a narrow FWHM in XRD rocking curves. We have analyzed the structural characteristics of MnTe films grown in this window. Figure 3(a) shows reciprocal space mapping (RSM) of Sample D around the asymmetric InP (224) diffraction. A mosaic of MnTe ($10\bar{1}5$) diffraction is also captured in addition to the anticipated InP (224) diffraction, where the in-plane component ($Q_x$) of MnTe matches closely that of InP. Using the center of the mosaic pattern, the in-plane lattice constant $a$ of MnTe is calculated to be 4.17 Å, which is slightly larger (< 0.8%) than the value reported in bulk crystals (4.15 Å),[41] and indicates that such 30-nm MnTe layer is nearly fully relaxed with minimal residual strain.

Cross-sectional TEM studies were carried out to characterize the structural properties of the MnTe films grown with conditions identical to Sample D but with a film thickness of ~ 100 nm. A low-magnification cross-sectional HAADF-SEM image along with the In, Mn, and Te energy dispersive spectroscopy mapping is shown in Figure S3. The sharp and clear transition from the In signal to the Te and Mn signals suggests an abrupt interface between the substrate and the epitaxial film. We also note that a thin amorphous oxide layer has formed on top of the sample, which is likely due to the natural oxidation of the film. High-resolution HAADF-STEM images are shown in Figure 3(b-c), further confirming the sharp interface between MnTe and InP. We see that the as-grown MnTe film is the α-polytype by superimposing the NiAs-type hexagonal crystal lattice on top of the HAADF-STEM image in Figure 3(c).[42] It is also worth noting that upon inspecting the entire FIB sample, no additional polytype or phases of MnTe are found, consistent with the XRD scans and again confirming the polytype purity and excellent quality of the MnTe film.

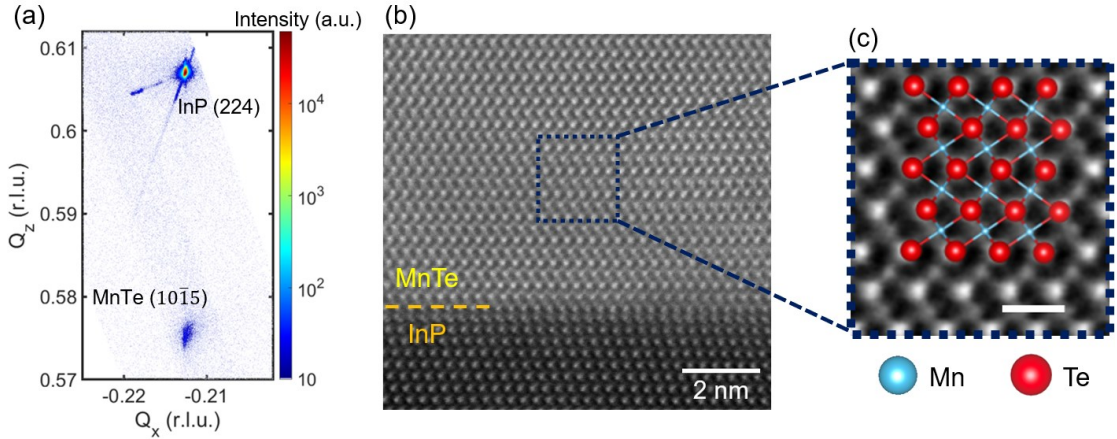

**Figure 3.** (a) XRD reciprocal space maps around the InP (224) diffractions. Both InP (224) and MnTe ($10\bar{1}5$) mosaics are labeled. (b) HAADF-STEM image showing the MnTe/InP interface. The yellow dashed line indicates the location of MnTe/InP interface. (c) High magnification HAADF-STEM image of the boxed region in (b), superimposed with the NiAs-type crystal structure created in VESTA. The scale bar in (c) is 0.5 nm.

3.3 Neutron diffraction

High angle neutron diffraction measurements were performed on 35 nm and 100 nm MnTe films which were grown with identical conditions to Sample D to determine if the magnetic structures of these films are consistent with bulk behavior. A distinct 0 0 0 1 peak, which is structurally forbidden, appears at low temperature, confirming that antiferromagnetic order comparable to bulk develops in both films. The temperature dependence of the integrated intensity of the antiferromagnetic peak is plotted in Fig. 4(a) and Fig. 4(b) for the 35 nm and 100 nm films, respectively, and follows the expected order parameter behavior. Fits of these data to a Brillouin function indicate that $T_N = 307 \pm 2$ K for the 35 nm film and $304 \pm 1$ K for the 100 nm film. These values are approximately equal to the bulk ordering temperature and indicate that the antiferromagnetic component film magnetism mimics that of bulk MnTe.

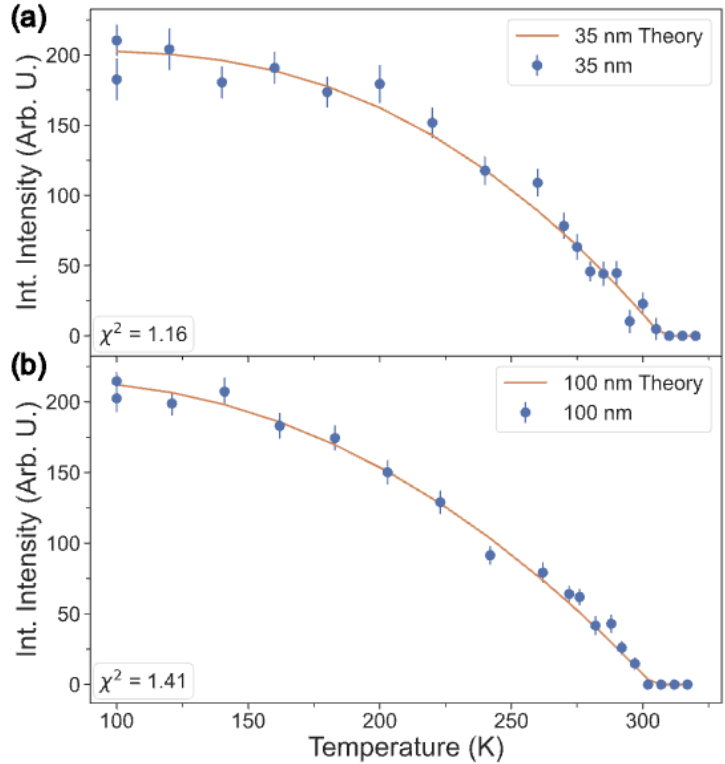

**Figure 4.** Integrated intensity of the 0 0 0 1 antiferromagnetic reflection as a function of temperature for (a) a 35 nm MnTe film and (b) a 100 nm MnTe film. The lines correspond to a fit to a Brillouin function.

3.4 Polarized neutron reflectometry

To explore the possibility of an additional ferromagnetic component of the MnTe magnetization, PNR measurements were taken at 20 K and 100 K, below the $T_N$ of MnTe, in a 1 T magnetic field directed within the film's surface. This field was expected to align any ferromagnetic (FM) regions in the MnTe into the plane of the film along the field direction. In this experiment, the 35 nm MnTe film is again grown with growth conditions identical to Sample D then capped with a 5-nm amorphous Te layer. Two models of the scattering length density profiles as a function of depth (Figs. 5(a) and 5(c)) generated best fits of equivalent quality, and so we could not rule out either as a potential solution. In the first model, referred to as the "bulk" model, the magnetization $\rho_M$ in the MnTe layer was uniform. Figure 5(a) shows the resulting $\rho$ and $\rho_M$ profiles for the two measurement temperatures, with the $\rho_M$ profile rescaled in Fig. 5(b) for clarity. Fit independently, the other model, referred to as the "interfacial magnetization" model, converged to a profile with all the same characteristics of the bulk model (within error), except $\rho_M$ in MnTe was decoupled from the thickness of the layer (Figs. 5(c) and 5(d)). A feature of both models is that the fitted nuclear $\rho$ is uniform through the entire MnTe layer thickness, and the values obtained in these fits are $0.440 \times 10^{-4}$ nm$^{-2}$ (bulk model) and $0.446 \times 10^{-4}$ nm$^{-2}$ (interfacial magnetization model). The 95% confidence intervals (CI) for these models $\rho$ are $0.421 - 0.457 \times 10^{-4}$ nm$^{-2}$ and $0.428 - 0.461 \times 10^{-6}$ nm$^{-2}$, for the bulk and interfacial magnetization models, respectively, which are slightly higher than the bulk $\rho$ value of MnTe of $0.379 \times 10^{-4}$ nm$^{-2}$. This result suggests that the structural quality of the film is comparable to that of bulk crystals, though slight Mn-deficiency or strain compression along the c-axis may give rise to a slight increase in the apparent density. The latter, however, is unlikely since the lattice parameters measured by XRD approximate those of bulk.

The depth profiles in Figure 5 were generated from the reflectivity data and fits for the 20 K measurements shown in Figure 6(a) (bulk model) and Figure 6(c) (interfacial magnetization model). It should be noted that the discontinuity in the data results from the different detector angles used during the time-of-flight measurements, with overlapping regions in the data along Q. The 100 K reflectivity data (Figures S4 – S7 in Supplemental Material) did not vary significantly from the 20 K data. The reflectivity fits corresponding to both models, shown as solid lines, are in good agreement with the data, and have similar goodness of fit ($\chi^2$), highlighting their equivalence as potential solutions.

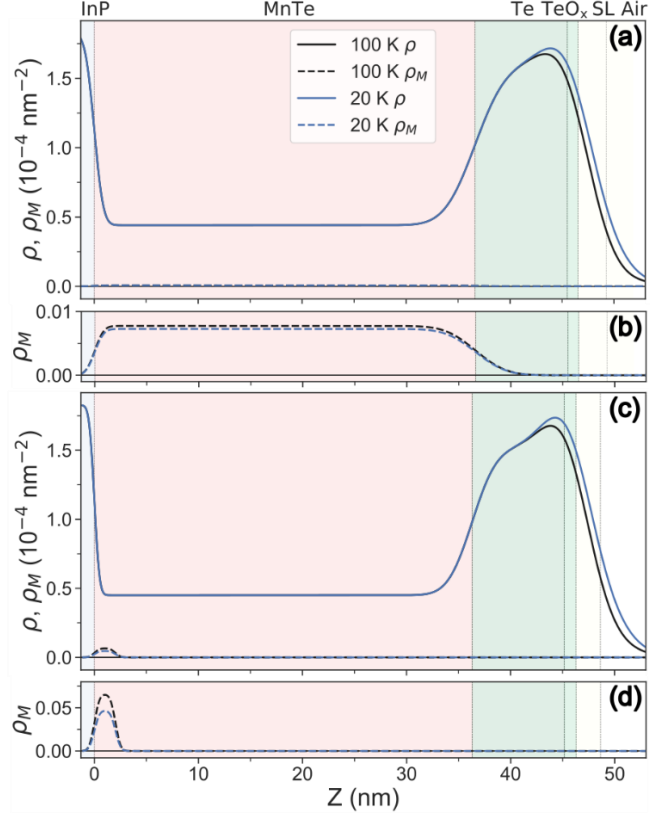

**Figure 5.** PNR $\rho$ and $\rho_M$ profiles calculated from reflectivity fits of Te/MnTe at 100 K (black) and 20 K (blue) from the bulk layer model (a, b) and the interface magnetism model (c, d). Solid lines represent $\rho$ and the dashed lines represent $\rho_M$ (a, c) and zoomed regions of $\rho_M$ (b, d) were used to illustrate the magnetic profiles. Background colors with corresponding labels above the plot highlight expected layer positions along the sample depth (Z). "SL" represents a surface layer of ice/condensation formed during sample cooling.

Spin asymmetry – defined as $(R^+ - R^-) / (R^+ + R^-)$ – is useful for highlighting the differences between the two polarization states that arise from the presence of in-plane magnetization. The spin asymmetry data and the corresponding fits for the bulk model (Figs. 5(a) and 5(b)) and interfacial magnetization model (Figs. 5(c) and 5(d)) are shown in Fig. 6(b) and Fig. 6(d), respectively. In general, the magnitude of the spin asymmetry is small, and the variations about zero are mostly within the level of uncertainty – indicative of a small magnetic contribution in the sample. While, qualitatively, the interfacial magnetization model seems to better capture the shape of the spin asymmetry, the large uncertainty above $Q \approx 0.25$ nm$^{-1}$ limits the determination of the true shape of the data – again suggesting equivalence in the models as potential solutions.

**Table 1.** $\rho_M$ Fit Values and Uncertainty for the Bulk and Interfacial Magnetization Models

| Model | T (K) | $\rho_M$ ( × 10$^{-4}$ nm$^{-2}$) | $M_{fit}$ (kA/m) | 95% CI ( × 10$^{-4}$ nm$^{-2}$) | $M_{max}$ (kA/m) |
|---|---|---|---|---|---|
| Bulk | 100 | 0.0080 | 2.8 | 0.0033 – 0.0123 | 4.31 |
|  | 20 | 0.0074 | 2.6 | 0.0042 – 0.0105 | 3.68 |
| Interfacial Mag. | 100 | 0.134 | 48 | 0.03 – 0.36 | 130 |
|  | 20 | 0.094 | 33 | 0.02 – 0.25 | 88 |

Table 1 summarizes the 100 K and 20 K fit values for $\rho_M$ for the bulk and interfacial magnetization models and their 95% confidence interval (CI), calculated from uncertainty analysis with a Markov Chain Monte Carlo method (Bayesian approach).[43] The 95% CI does not contain zero for either model, suggesting a finite magnetization persists in both cases. For the bulk model (Fig. 5(b)), a uniform, vanishingly small $\rho_M$ resulted from fits of both the 100 K and 20 K data. In the interfacial magnetization model (Fig. 5(d)), $\rho_M$ is significantly larger, but is confined to a thickness of 1.38 nm ± 0.87 nm at the MnTe/InP interface (for both 100 K and 20 K). We find no evidence for a transitional interface region, as models which allowed for an interface region with an $\rho$ distinct from that of the rest of the MnTe film did not improve the fit (Fig. S10 and Fig. S11 in Supplemental Material). PNR is therefore consistent with a uniform structure throughout the entire film depth.

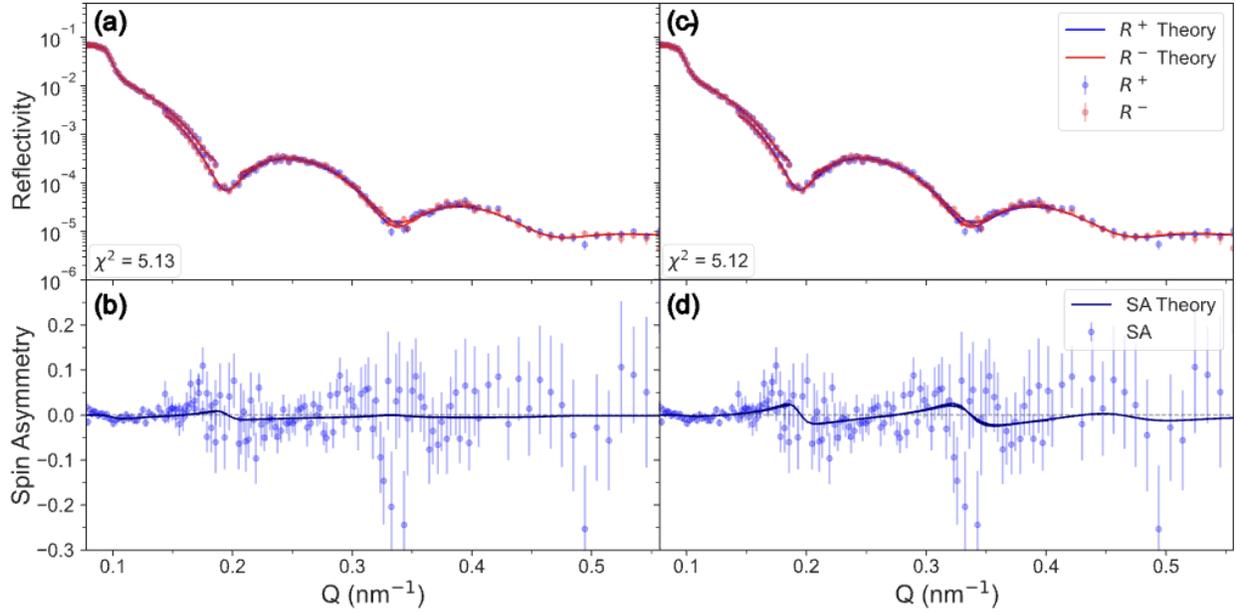

**Figure 6.** PNR data (points) and theoretical curves (solid lines) calculated from fits using the bulk layer (a, b) and interfacial magnetization models (c, d) for measurements at 20 K on Te (10 nm)/MnTe (35 nm)/ InP films. The reflectivity data and fits (a, c) were used to calculate the spin asymmetry (b, d) defined as $(R^+ - R^-) / (R^+ + R^-)$. Error bars on the reflectivity data and calculated spin asymmetry represent the standard deviation.

To explore the most extreme case, a maximum, laterally averaged magnetization, $M_{max}$, was calculated from the upper 95% CI bound using the conversion constant $c = 2.853 \times 10^{-7}$ $nm^{-2}\,kA^{-1}\,m$ (Table 1). For the bulk model, the $M_{max} \approx 4$ kA/m is effectively negligible, but would be consistent with a very small number of FM impurities and/or clusters uniformly distributed throughout the bulk of the MnTe layer. For the interfacial magnetization model, it should be noted that the majority of the film has zero magnetization, and the values indicated in the table represent only the thin sublayer at the MnTe/InP interface with $M_{max} \approx 90 - 130$ kA/m. Such a localized magnetization could be caused by a small amount of diffusion and/or strain isolated in the thin interfacial layer, though the uniformity of the nuclear ρ profile (Fig. 5(c)) does not support this interpretation. Similarly, a slight increase in Mn content has been attributed to FM in MnTe,[33] but this does not explain FM in either of our films since we observed a uniform increase in ρ in the MnTe layer from the expected value, consistent with a Mn deficiency.

While the above analysis focuses on the upper limit of the MnTe magnetization established by PNR, the best-fit values, $M_{fit}$, for both the bulk and interfacial magnetization models (Table 1) are a factor of two, or more, than $M_{max}$. In addition, the average magnetization $M_{fit}$ for the bulk model is substantially less than the magnetization measured by neutron reflectivity on previously reported MnTe films,[19, 33] which ranges from 10 – 50 kA/m. Any ferromagnetic component of the magnetization through the MnTe film is either trace or confined to a very narrow interfacial region. Both possibilities indicate that the high quality MnTe film is bulk-like in character with good stoichiometry and very limited magnetic clustering or local defects.

3.5 Angle Resolved Photoemission Spectroscopy

Theoretical calculations and experimental ARPES measurements on MnTe bulk crystals and thin films have shown altermagnetic lifting of Kramer's spin degeneracy of bulk bands.[22] At positions with high symmetry in momentum space such as the nodal planes $\Gamma - M - K$ and $\Gamma - A - H - K$, a 'weak' altermagnetic band splitting is observed, arising from the influence of spin-orbit coupling. Away from the nodal planes, the bands exhibit a large spin splitting due to nonrelativistic 'strong' altermagnetic lifting of Kramers degeneracy. To compare the band structure of our MnTe thin films with prior results, we carried out *in vacuo* angle resolved photoemission spectroscopy ARPES measurements of 35 nm thick MnTe thin films using a Helium lamp with 21.2 eV photon energy. Prior ARPES experiments on MBE grown MnTe thin films with similar photon energy show that the $k_z$ value is in the high symmetry nodal plane $\Gamma - M - K$.[22] Below the Neel temperature (307 K, as shown by neutron diffraction), we expect to observe a weak altermagnetic band splitting at the top of the valence band. Figure 7 (a) shows the sixfold symmetric constant energy map of the thin film integrated over 70 meV binding energy from the Fermi level. From the hexagonal symmetry of the map, we can conclude that the sample has equal populations of domains with Néel vector easy axis pointing along three equivalent crystallographic directions ($\Gamma - M$ directions). We also map the band spectrum along the $\Gamma - K$ direction at 298 K (Fig. 7(b)) and at 77 K (Fig. 7(c)). The high spectral intensity bands observed in the measurements (also see supplementary note on ARPES) are similar to those seen in prior studies[22] of MnTe films at $k_z = 0$. Comparing the ARPES data at 77 K and 298 K, we note that in the former, the top of the valence band (Fig. 7(c)) with high spectral intensity (brown) is different from the low intensity spectral bands (green) just below it; in the latter (298 K), this difference is absent. We associate this temperature dependence of the valence band splitting to the altermagnetic phase transition in our MnTe films. As expected, the bands become sharper when measured at lower temperatures (see Fig. 7 (b-c)) due to smaller thermal broadening. However, at the lowest measured temperature (77 K), we still do not unambiguously observe the predicted weak altermagnetic splitting of the bands along $\Gamma - K$. We anticipate that lower temperatures are needed to measure this splitting as shown in prior work.[26] Further, to confirm the much larger 'strong' altermagnetic band splitting, we would need to carry out future ARPES measurements under appropriate conditions that access momentum space away from the nodal planes.

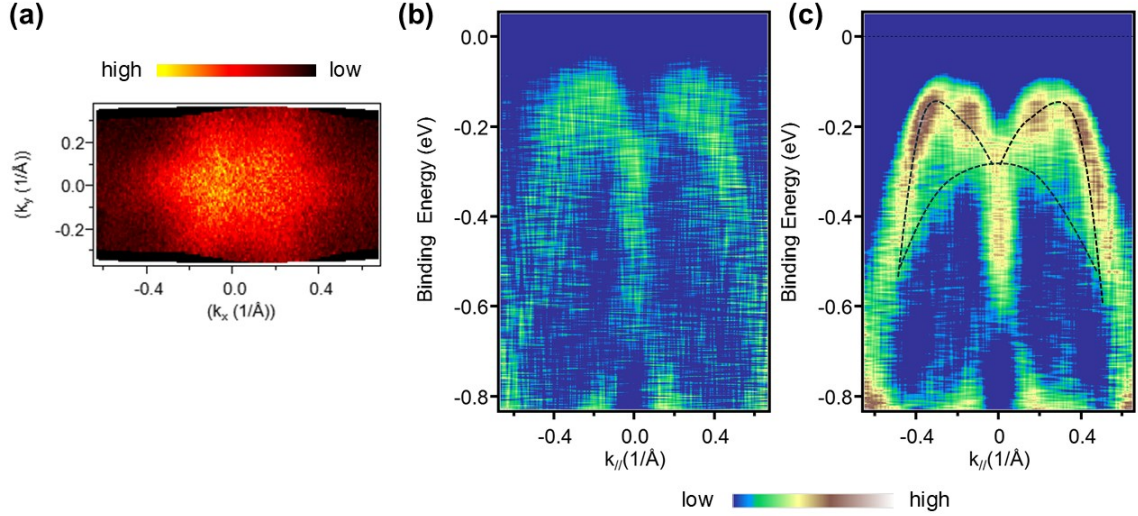

**Figure 7.** ARPES data on 35 nm thick MnTe film: (a) Constant energy map of the film integrated from 70 meV binding energy to the Fermi surface. The data are measured at 77 K. (b-c) Band spectrum measured along $\Gamma - K$ at (b) 298 K and (c) at 77 K. The black dashed lines are guides to the eye for the splitting in the valence band.

## 4. Summary

In conclusion, we have reported an in-depth study on the magnetic spin structure of α-MnTe films with exceptional quality. We first identified an optimized growth window of MnTe (film growth rate of 0.35Å/s, Te/Mn flux ratio of 3 – 4, substrate temperature of 250 – 400 °C) on InP substrates, which produces ultrahigh quality MnTe films including smooth surfaces with RMS roughness of ~ 1 nm and narrow FWHM of 0.1° and 0.3° at MnTe 0004 and $10\bar{1}5$ XRD rocking curves. A single-phase α-MnTe film was further confirmed with reciprocal space mapping and scanning transmission electron microscopy experiments. Using a 35 nm MnTe film grown with optimized conditions, neutron diffraction was conducted to confirm the antiferromagnet nature of the synthesized film, which has a Néel temperature of ~307 K. We further used polarized neutron reflectometry to investigate the possibility of a FM component in the MnTe magnetization, with the modeled $\rho_M$ indicating a vanishingly small average magnetization of 2.6 kA/m that is more than a factor of two to four smaller than previously reported values. The net magnetization is either attributed to a very small number of impurities in the MnTe film or confined to a narrow MnTe/InP interfacial region. These results thus highlight the ultrahigh quality in the synthesized MnTe film with a nearly ideal stoichiometric order.

DISCLAIMER

We identify certain commercial equipment, instruments, and materials in this article to specify adequately the experimental procedures. In no case does such identification imply recommendation or endorsement by the National Institute of Standards and Technology nor does it imply that the materials or equipment identified are necessarily the best available for the purpose.

ACKNOWLEDGMENT

This research was conducted at the Pennsylvania State University Two-Dimensional Crystal Consortium – Materials Innovation Platform which is supported by NSF cooperative agreement DMR-2039351. The authors appreciate the use of the Penn State Materials Characterization Lab.

AUTHOR INFORMATION

Corresponding Authors

**Qihua Zhang –** [qzz5173@psu.edu](mailto:qzz5173@psu.edu)

**Stephanie Law –** [sal6149@psu.edu](mailto:sal6149@psu.edu)

**Data availability statement**

All data of this study is available under the following link for the review process, which will be converted into an open-access link to the data with separate DOI in ScholarSphere upon publication: [https://data.2dccmip.org/8LP0hbN3XhSJ](https://data.2dccmip.org/8LP0hbN3XhSJ).

For Table of Content Only

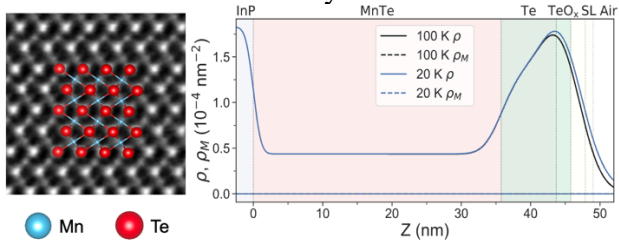

# Supporting Information

# Examining the Spin Structure of Altermagnetic Candidate MnTe Grown with Near Ideal Stoichiometry

Qihua Zhang[1*], Christopher J. Jensen[2], Alexander J. Grutter[2], Sandra Santhosh[3], William D. Ratcliff[2,4,5], Julie A. Borchers[2], Thomas W. Heitmann[6], Narendirakumar Narayanan[6], Timothy R. Charlton[7], Mingyu Yu[8], Ke Wang[9], Wesley Auker[9], Nitin Samarth[1,3,9,10,11], Stephanie Law[1,3,9,10,12]*

[1]Two Dimensional Crystal Consortium Materials Innovation Platform, The Pennsylvania State University, University Park, Pennsylvania 16802, USA

[2]NIST Center for Neutron Research, NCNR, National Institute of Standards and Technology, Gaithersburg, Maryland 20899, USA

[3]Department of Physics, The Pennsylvania State University, University Park, Pennsylvania 16802, USA

[4]Department of Physics, University of Maryland, College Park, Maryland 20742, USA

[5]Department of Materials Science and Engineering, University of Maryland, College Park, Maryland 20742, USA

[6]University of Missouri Research Reactor (MURR), University of Missouri, Columbia, Missouri 65211, USA

[7]Neutron Scattering Division, Oak Ridge National Laboratory, Oak Ridge, Tennessee 37831, USA

[8]Department of Materials Science and Engineering, University of Delaware, 201 Dupont Hall, 127 The Green, Newark, Delaware 19716, USA

[9]Materials Research Institute, The Pennsylvania State University, University Park, Pennsylvania 16802 USA

[10]Department of Materials Science and Engineering, The Pennsylvania State University, University Park, Pennsylvania 16802 USA

[11]Argonne National Lab, Chicago, Illinois 60439 USA

[12]Institute of Energy and the Environment, The Pennsylvania State University, University Park, Pennsylvania 16802 USA

CORRESPONDING AUTHORS

Qihua Zhang – qzz5173@psu.edu

Stephanie Law – sal6149@psu.edu

**Table S1.** Growth parameters (Mn flux, Te flux, Growth Temperature, Te/Mn FR) of the MnTe layer in Samples A-L. The RMS roughness of the 2 μm × 2 μm AFM images are also included.

| Sample No. | Mn Flux (cm$^{-2}$/s) | Te Flux (cm$^{-2}$/s) | Growth Temperature (°C) | Te/Mn FR | RMS roughness (nm) | MnTe (0004) 2θ position |
|---|---|---|---|---|---|---|
| A | $7.6 \times 10^{13}$ | $2.3 \times 10^{14}$ | 200 | 3 | 0.8 | 54.70° |
| B | $7.6 \times 10^{13}$ | $2.3 \times 10^{14}$ | 250 | 3 | 0.9 | 54.76° |
| C | $7.6 \times 10^{13}$ | $2.3 \times 10^{14}$ | 300 | 3 | 1.0 | 54.80° |
| D | $7.6 \times 10^{13}$ | $2.3 \times 10^{14}$ | 340 | 3 | 1.4 | 54.80° |
| E | $7.6 \times 10^{13}$ | $2.3 \times 10^{14}$ | 400 | 3 | 1.9 | 54.81° |
| F | $7.6 \times 10^{13}$ | $2.3 \times 10^{14}$ | 500 | 3 | 3 | 54.86° |
| G | $7.6 \times 10^{13}$ | $7.6 \times 10^{13}$ | 340 | 1 | 0.8 | 54.87° |
| H | $7.6 \times 10^{13}$ | $1.5 \times 10^{14}$ | 340 | 2 | 2.0 | 54.83° |
| J | $7.6 \times 10^{13}$ | $3.1 \times 10^{14}$ | 340 | 4 | 1.2 | 54.79° |

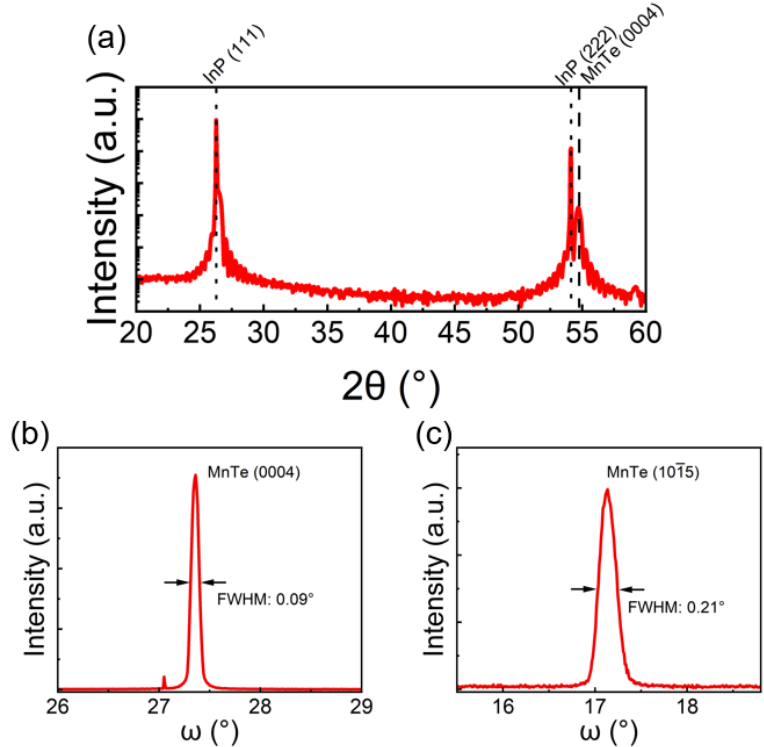

**Figure S1.** High resolution XRD characterization of Sample A. (a) XRD 2θ-ω coupled scan. The InP diffraction peaks are marked by dotted lines; MnTe diffraction peaks are marked by dashed lines. (b) ω-scan of the MnTe 0004 reflection. (c) ω-scan of the MnTe $10\bar{1}5$ reflection.

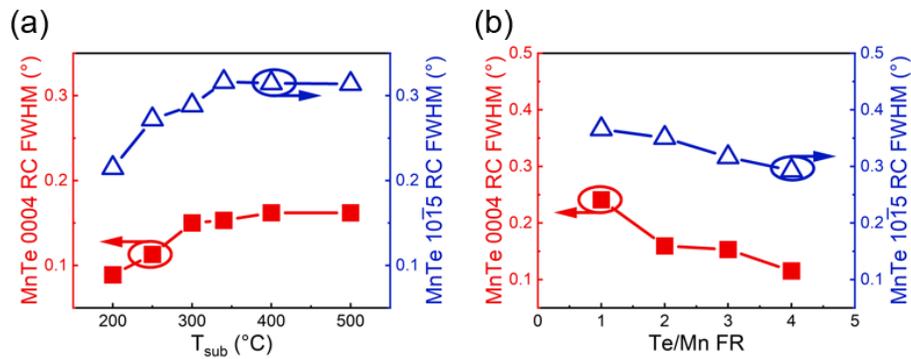

**Figure S2.** (a) FWHM of the MnTe 0004 (in red solid squares) and $10\bar{1}5$ (in blue open triangles) reflections from XRD rocking curves as a function of growth temperature. (b) FWHM of the MnTe 0004 (in red solid squares) and $10\bar{1}5$ (in blue open triangles) reflections from XRD rocking curves as a function of Te/Mn flux ratio.

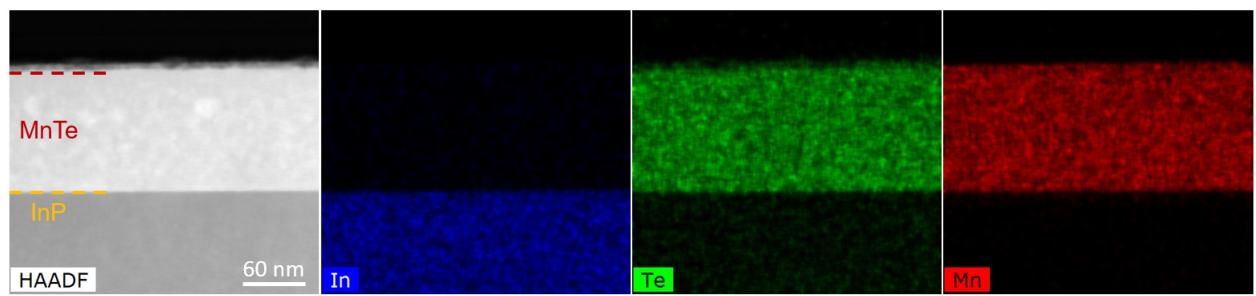

**Figure S3.** Cross-sectional HAADF-STEM image (left) and EDS mapping of In, Te, and Mn. The dashed lines in the HAADF image separate the MnTe film and InP substrate.

## Polarized Neutron Reflectometry (PNR)

**100 K –Reflectivity and Spin Asymmetry**

The fits for the reflectivity measured at 100 K of the Te (10 nm)/ MnTe (35 nm)/ InP sample are shown in Fig. S4 (bulk model) and S6 (interfacial magnetization model). These fits, and the fits of the 20 K reflectivity data (Fig. 6), were generated from the nuclear and magnetic scattering length densities ($\rho$ and $\rho_M$) shown in Fig. 5 of the main text. Fig. S5 (bulk model) and S7 (interfacial magnetization model) are the calculated spin asymmetry $[(R^+ - R^-) / (R^+ + R^-)]$ for the data and fits.

*Bulk Model*

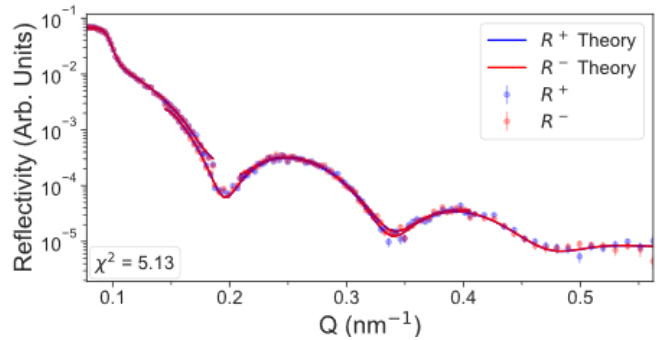

**Figure S4.** PNR data (points) and theoretical fits (solid lines) calculated using the bulk layer model for measurements at 100 K on Te (10 nm)/MnTe (35 nm)/InP films).

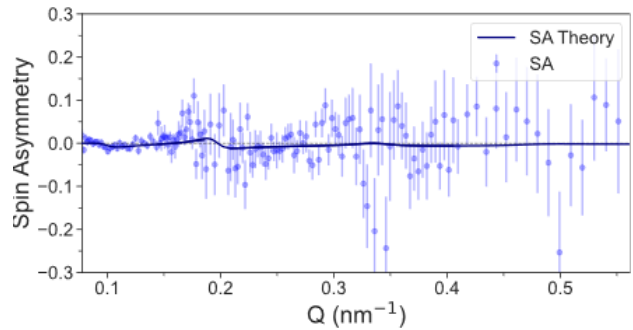

**Figure S5.** Spin asymmetry calculated from the data (points) and theoretical fits (solid lines) using the bulk model for the PNR measurements taken at 100 K on Te (10 nm)/MnTe (35 nm)/InP films. Spin asymmetry is defined as $(R^+ - R^-) / (R^+ + R^-)$.

*Interfacial Magnetization Model*

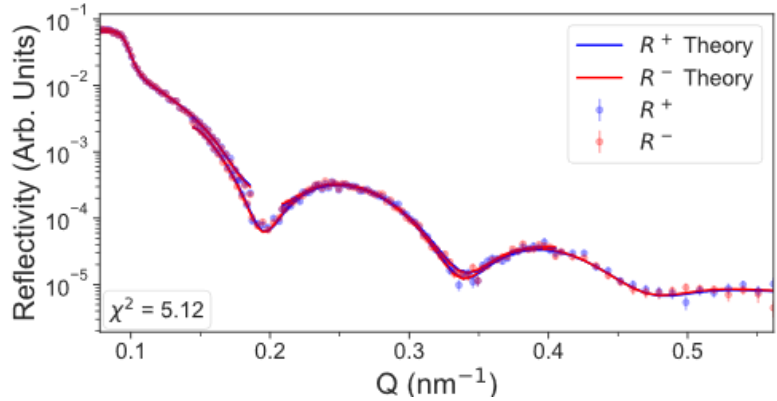

**Figure S6.** PNR data (points) and theoretical fits (solid lines) calculated using the interfacial magnetization model for measurements at 100 K on Te (10 nm)/MnTe (35 nm)/InP films.

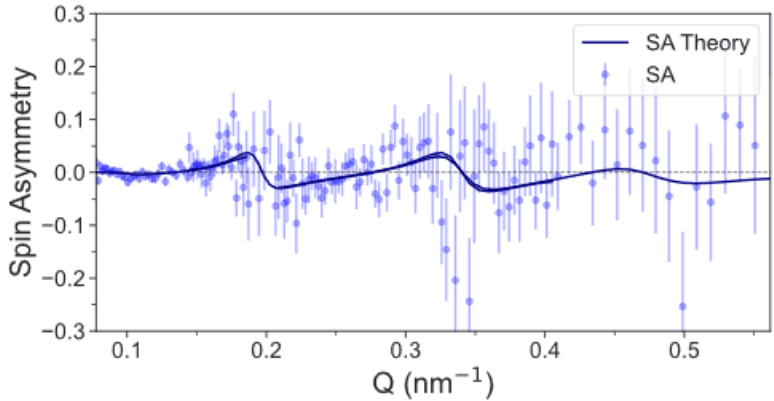

**Figure S7.** Spin asymmetry calculated from the data (points) and theoretical fits (solid lines) using the interfacial magnetization model for the PNR measurements taken at 100 K on a Te (10 nm)/MnTe (35 nm)/InP film. Spin asymmetry is defined as $(R^+ - R^-) / (R^+ + R^-)$.

**Further Explanation on Fitting Process**

Polarized neutron reflectivity data were fit using a series of models, starting from a simple structure with uniform $\rho$ for each layer in the structure and no $\rho_M$ (i.e., no magnetization allowed to be fit in MnTe). From this, a series of more complex models were considered to identify the best fit model(s) – as shown in the main text Fig. 5 – and those that were believed to be under or over parameterized. To determine the quality of each fit, the goodness of fit ($\chi^2$) was considered, but also the qualitative match between the shape of the calculated theoretical reflectivities and the spin asymmetries and data. The latter is especially relevant for this study, as extracting the magnitude and distribution of the magnetization is critical for understanding altermagnetic characteristics of MnTe. Below we discuss the quality of the fits in the context of two models that were under-parameterized and over-parameterized, referred to as "non-magnetized MnTe" and "interfacial MnTe layer", respectively.

*Under Parameterized Model – Non-Magnetized MnTe*

The best-fit $\rho$ and $\rho_M$ profiles from the non-magnetized MnTe model are shown in Fig. S8 for the 100 K and 20 K measurements of the Te (10 nm)/ MnTe (35 nm)/ InP sample. A uniform $\rho$ was used for each layer in the sample, and the interfacial widths of each layer were fit, except for that of TeO$_x$, which was linked to the interfacial width of the Te layer. By way of definition, the interfacial width is the full width of the error function used to describe the layer interfaces assuming Gaussian roughness. The origin of this interfacial mixing could be conformal roughness, thickness variations and/or chemical interdiffusion averaged across the entire 1 cm$^2$ sample plane. The width of the InP/MnTe interface obtained from all fits described here was $0.5 \pm 0.3$ nm, and the width of the MnTe/Te interface was $2.1 \pm 0.4$ nm. These values confirm that the MnTe film is structurally uniform through the majority of its depth.

The fits of the 100 K and 20 K data were conducted simultaneously, with all layers' structural properties assumed to be equal between 100 K and 20 K, except for the surface layer ("SL" in Fig. S8). As described in the main text, this surface layer is likely a result of ice or condensate accumulation on the surface of the film while cooling, and the increased thickness at 20 K is consistent with this explanation. The MnTe was fit with a $\rho = 0.439 \times 10^{-4}$ nm$^{-2}$, which is higher than the expected value for bulk but about the same as the bulk and interfacial magnetization models in the main text. The $\rho_M$ for all layers were forced to be zero, which is representative of a film with a MnTe layer that is devoid of any magnetization.

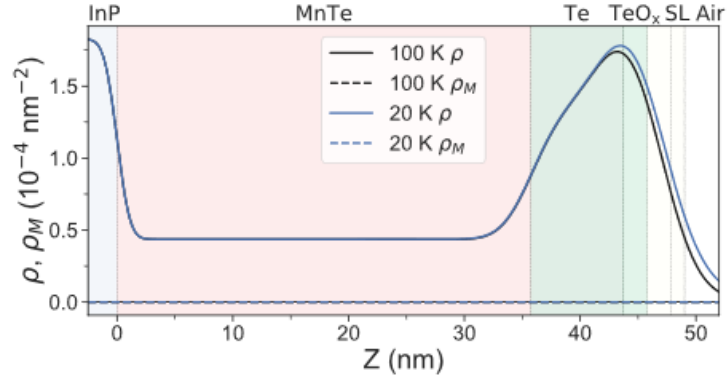

**Figure S8.** PNR $\rho$ and $\rho_M$ profiles calculated from fits of data collected for the Te (10 nm)/MnTe (35 nm)/InP film at 100 K (black) and 20 K (blue) using the non-magnetized MnTe model. Solid lines represent $\rho$ and the dashed lines (hidden by origin) represent $\rho_M$. Background colors with corresponding labels above the plot were used to highlight the expected layer positions along the depth (Z) of the sample. Note that "SL" represents a likely surface layer of ice/condensation that formed during sample cooling.

The 20 K (Fig. S9(a)) and 100 K (Fig. S9(c)) reflectivity are both fit well by this model, but with a slightly higher value of $\chi^2$ compared to the best fits included in the main text (Fig. 6). When spin asymmetry is calculated (Fig. S9(b) and S9(d)), the theoretical curve is a flat line at zero since there is no splitting in the fitted $R^+$ and $R^-$ reflectivities. This clearly does not match the fluctuations in the spin asymmetry observed in the data and contributes to the slight increase in $\chi^2$, so this model was considered under parameterized. Additionally, it justifies that a magnetic layer is needed in the model to produce the appropriate splitting measured between the $R^+$ and $R^-$.

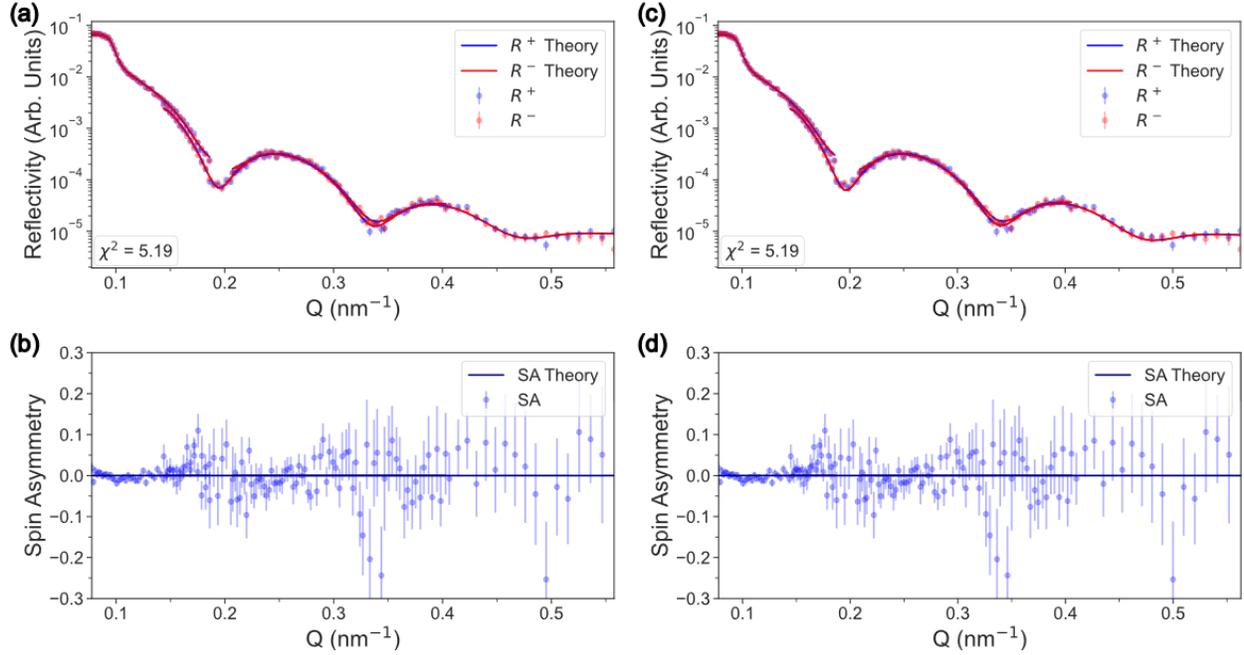

**Figure S9.** PNR data (points) and theoretical fits (solid lines) calculated using the non-magnetized MnTe model for measurements at (a) 20 K and (c) 100 K on Te (10 nm)/MnTe (35 nm)/InP films). Spin asymmetry was calculated from these data (points) and theoretical fits (solid lines) for the (b) 20 K and (d) 100 K on a Te (10 nm)/MnTe (35 nm)/InP film. Spin asymmetry is defined as $(R^+ - R^-)/(R^+ + R^-)$.

*Over Parameterized Model – Interfacial MnTe Layer*

The interfacial MnTe layer model's calculated $\rho$ and $\rho_M$ profiles are shown in Fig. S10 for the 100 K and 20 K measurements of the Te (10 nm)/ MnTe (35 nm)/ InP sample. This model introduced a thin MnTe$_{int}$ layer at the MnTe/InP interface that may be fit with different $\rho$ and $\rho_M$ values than those of the MnTe layer. By introducing MnTe$_{int}$, this model is very similar to the interfacial magnetization model chosen as one of the best fits (Fig. 5(c)) but differs since $\rho$ was allowed to vary in this interface region. During the fitting process, the fit values for each layer were constrained to be the same between the 100 K and 20 K (assuming no large structural changes), except for the values for $\rho_M$ in MnTe and the structural parameters for the surface layer ("SL" Fig. S10). The increase in the surface layer thickness at 20 K is again consistent with the ice or condensate accumulation during the cooling process. MnTe in this model resulted in $\rho = 0.443 \times 10^{-4}$ nm$^{-2}$, which is comparable to that obtained from the other models. The MnTe$_{int}$ $\rho = 0.119 \times 10^{-4}$ nm$^{-2}$ with a thickness of 1.0 nm between the InP and MnTe layers. This reduction in $\rho$ could correspond to a transition layer being formed during growth with lower density or a higher percentage of Mn. Magnetization in this model is confined to MnTe$_{int}$, with $\rho_M = 0.128 \times 10^{-4}$ nm$^{-2}$ at 100 K and $\rho_M = 0.095 \times 10^{-4}$ nm$^{-2}$ at 20 K, corresponding to M$_{fit}$ of 45 kA/m and 33 kA/m, respectively. These values are comparable to the magnetization obtained using the interfacial magnetization model (Table 1) and are consistent with similar findings of ferromagnetism being present in Mn-rich MnTe films.[1]

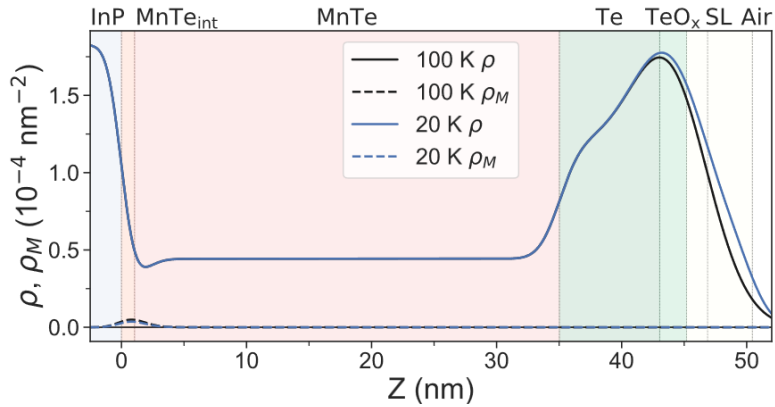

**Figure S10.** PNR $\rho$ and $\rho_M$ profiles calculated from fits of data collected for the Te (10 nm)/MnTe (35 nm)/InP film at 100 K (black) and 20 K (blue) using the interfacial MnTe layer model. Solid lines represent $\rho$ and the dashed lines represent $\rho_M$. Background colors with corresponding labels above the plot were used to highlight the expected layer positions along the depth (Z) of the sample. Note that "SL" represents a likely surface layer of ice/condensation that formed during sample cooling.

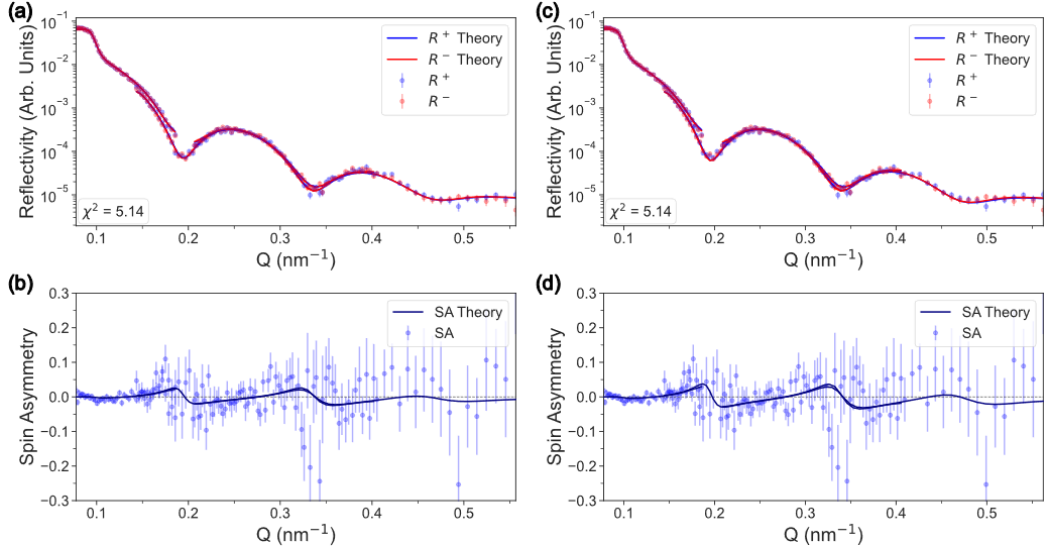

**Figure S11.** PNR data (points) and theoretical fits (solid lines) calculated using the interfacial MnTe layer model for measurements at (a) 20 K and (c) 100 K on Te (10 nm)/MnTe (35 nm)/InP films). Spin asymmetry was calculated from these data (points) and theoretical fits (solid lines) for the (b) 20 K and (d) 100 K on a Te (10 nm)/MnTe (35 nm)/InP film. Spin asymmetry is defined as $(R^+ - R^-) / (R^+ + R^-)$.

When compared to the interfacial magnetization model (Fig. 5(c), 5(d) and Fig. S6 and S7), the reflectivity and spin asymmetry in Fig. S11 are very similar, and $\chi^2$ is almost identical. Because of this, it is hard to justify this model over the interfacial magnetization model, and the addition of ρ variation in the MnTe$_{int}$ layer is not critical to the overall fit. We consider this model to be over-parameterized though we cannot rule out the possibility of a thin interfacial, transition layer with reduced structural density and isolated ferromagnetism.